\documentclass[12pt,a4paper]{article}

\usepackage{amsmath,amssymb}
\usepackage{bm}
\usepackage{graphicx}
\usepackage{ascmac}
\usepackage{wrapfig}
\usepackage{braket}
\usepackage{color}
\usepackage{mathrsfs}
\usepackage[english]{babel}
\usepackage{inconsolata}
\usepackage{cases}
\usepackage{tikz}
\usepackage{cite}
\usepackage{setspace}
\usepackage[top=3cm, bottom=1.8cm, left=2.3cm, right=2.3cm, includefoot]{geometry}
\usepackage{hyperref}

\newcommand{\beq}{\begin{equation}}
\newcommand{\eeq}{\end{equation}}
\newcommand{\lr}[1]{\left(#1\right)}
\newcommand{\lrk}[1]{\left[#1\right]}

\makeatletter
\@addtoreset{equation}{section}

\makeatother

\begin{document}

\begin{titlepage}

\pagenumbering{gobble}

\begin{flushright}
\rm TIT/HEP-691 \\ July, 2022
\end{flushright}

\renewcommand{\thefootnote}{*}

\vspace{0.2in}
\begin{center}
\Large{\bf Quasi-normal modes for the D3-branes \\ and Exact WKB analysis}
\end{center}
\vspace{0.2in}
\begin{center}
\large Keita Imaizumi\footnote{E-mail: k.imaizumi@th.phys.titech.ac.jp}
\end{center}

\begin{center}{\it
Department of Physics,
\par
Tokyo Institute of Technology
\par
Tokyo, 152-8551, Japan
}
\end{center}
\vspace{0.2in}
\begin{abstract}
\begin{spacing}{1.0}
{\footnotesize  
We demonstrate how the Exact WKB analysis works in the study of the quasi-normal modes (QNMs). We apply the Exact WKB analysis to a massless scalar perturbation to the D3-brane metric as a concrete example. The Exact WKB analysis provides an exact condition for the QNMs. We numerically check our exact condition by using the Borel-Pad\'{e} approximation. We also present an analytic form and an asymptotic behavior of the QNMs.}
\end{spacing}
\end{abstract}

\end{titlepage}

\newpage

\pagenumbering{arabic}
\setcounter{page}{1}

\renewcommand{\thefootnote}{\arabic{footnote}}
\setcounter{footnote}{0}

\section{Introduction}
\label{sec:intro}
The quasi-normal modes (QNMs) of a wave are the complex resonant frequencies with damped part. One of well-studied examples appearing the QNMs is the gravitational waves in the ring-down phase of binary black holes merger, firstly observed by \cite{1602.03837}. The gravitational waves in the ring-down phase can be described as the linear perturbation to the metrics. Moreover, the symmetries of the background spacetimes may reduce the gravitational linear perturbation to the scalar or the vector perturbation. 
\par 
The QNMs also appear in the string theory. The dilaton field propagating in the D3-branes background can be regarded as a scalar perturbation to the metric. It is checked that the QNMs computed by the world volume theory and the supergravity description of the D3-branes are consistent with each other \cite{hep-th/0208063}. Another important example is the AdS/CFT correspondence \cite{hep-th/9711200, hep-th/9802150}. Under the AdS/CFT correspondence, the perturbation to the black hole in the AdS side corresponds to the perturbation to the thermal equilibrium state of the CFT. Then the imaginary part of the lowest mode of the QNMs in the AdS side is equivalent to the timescale that the CFT approaches to the thermal equilibrium \cite{hep-th/9909056}. 
\par
Mathematically, the QNMs are complex resonant eigenvalues of second-order ordinary differential equations. The eigenvalue problem of the QNMs can be numerically solved by the Leaver's method \cite{Leaver}. If the absolute value of the QNMs is large, the eigenvalue problem can be analytically solved by the WKB approximation \cite{classicalWKB}, which is the leading order approximation of the all-order WKB method. In the WKB approximation, the QNMs are determined by the Bohr-Sommerfeld condition for the ``potential barrier'' of the differential equations. There are also some higher order WKB formulae for the QNMs (e.g.\cite{WKBapp, WKBapp2, 1704.00361, 1904.10333}).
\par
Beyond the WKB approximation, it is recently conjectured that the QNMs for the 4-dimensional Schwarzschild black hole and the Kerr black hole are determined by the all-order Bohr-Sommerfeld condition \cite{2006.06111}, which is the extension of the Bohr-Sommerfeld condition in the WKB approximation to the all-order WKB method version. The authors numerically solved the all-order Bohr-Sommerfeld condition and demonstrated that the conjecture is correct, even if the absolute value of the QNMs is small. The authors also pointed out that the differential equations of the eigenvalue problem coincide with the quantum Seiberg-Witten curves for 4-dimensional  $\mathcal{N} = 2\ SU(2)$ supersymmetric gauge theories \cite{1705.09120}. This identity enables us to utilize gauge theory techniques to calculate the QNMs. The quantum Seiberg-Witten/QNMs dictionary is extended to other cases including D3-branes, fuzzballs and exotic compact objects later \cite{2105.04245, 2109.09804}. The differential equations of the QNMs eigenvalue problem also coincide with the semiclassical limit of the differential equations for the conformal block with a degenerate field insertion in two-dimensional CFT \cite{2105.04483, 2201.04491, 2206.09437}. The connection problem in the CFT side provides the conditions satisfied by the QNMs, which can be identified with the all-order Bohr-Sommerfeld conditions in \cite{2006.06111, 2105.04245, 2109.09804} via the AGT correspondence \cite{0906.3219}.
\par
Aim of the present paper is to propose the Exact WKB analysis \cite {Vor, DP, Esemi, cr1}, which is a non-perturbative formulation of the all-order WKB method, as a new tool to study the QNMs. In this paper, we apply the Exact WKB analysis to a massless scalar perturbation in the D3-branes background, whose QNMs can be numerically calculated by the all-order Bohr-Sommerfeld condition \cite{2105.04245, 2109.09804} and the ODE/IM method \cite{2112.11434}. The Exact WKB analysis provides an exact condition, which asymptotic expansion produces the all-order Bohr-Sommerfeld condition. The exact condition determines the QNMs non-perturbatively, while the all-order Bohr-Sommerfeld condition determines them perturbatively. We also make a comment on applicability of the Exact WKB analysis.
\par
This paper is organized as follows. In section \ref{sec:EWKB}, we present the basics of the exact WKB analysis. We explain the Borel resummation of the WKB solutions and its property of the analytic continuation. In section \ref{sec:CPQNM}, we solve a connection problem of the WKB solutions for the case of the D3-brane metric in practice. We will show that the connection formula and the boundary conditions produce an exact condition for the QNMs. We also precent some numerical and analytic computations of the QNMs by using our exact condition.

\section{Exact WKB analysis of the second-order ODE}
\label{sec:EWKB}
Throughout this paper, we consider the following second-order differential equation,
\beq
\label{eq:EWKBdeq}
\lrk{-\eta^2\frac{d^2}{dz^2} + Q_0\lr{z} + \eta^2Q_2\lr{z}}\psi\lr{z} = 0,
\eeq
where
\beq
\label{eq:potentials}
Q_0\lr{z}=-\frac{\omega^2L^2\lr{z^4 + 1} - z^2\lr{l+2}^2}{z^4}, \ \ Q_2\lr{z} = -\frac{1}{4z^2},
\eeq
$z$ is a complex coordinate, $\eta$ and $\omega$ are complex parameters, and $L$, $l$ are real. 
\par
At $\eta = 1$ and under the rescaling of the coordinate $zL = r$, the differential equation (\ref{eq:EWKBdeq}) on the real-positive axis agrees with the radial direction of the E.O.M. for a massless scalar field in the D3-brane metric \cite{2105.04245}. In the context of the D3-brane picture, $L$ is the AdS-radius, $l \in \mathbb{Z}_{\geq 0}$ is the orbital angular momentum of the scalar field, and $\omega$ is the frequency of the scalar field.
\par
The all-order WKB method produces an asymptotic expansion in $\eta$ of the solution to (\ref{eq:EWKBdeq}),
\beq
\label{eq:AOWKBsol}
 \psi_a(z) = \exp(\frac{i}{\eta}\int_{a}^{z}\sum_{n=0}^{\infty}\eta^np_{n}(z')dz'), 
\eeq
where $a$ is an arbitrary point controlling the normalization of the solution. The solution (\ref{eq:AOWKBsol}) is called the WKB solution. By substituting the WKB solution into (\ref{eq:EWKBdeq}), we obtain $p_{n}(z)$ recursively. Especially we find $p_0(z) = \sqrt{-Q_0(z)}$. Moreover, dividing $\sum_{n=0}^{\infty}\eta^np_{n}(z)$ into the even power part and the odd power part as
\beq
 \sum_{n=0}^{\infty}\eta^np_{n}(z) = \sum_{n=0}^{\infty}\eta^{2n}p_{2n}(z) + \sum_{n=0}^{\infty}\eta^{2n + 1}p_{2n + 1}(z) = P_{\mathrm{even}}(z) + P_{\mathrm{odd}}(z),
\eeq
we find the following relation
\beq
 P_{\mathrm{odd}}(z) = -\frac{1}{2}\frac{\partial}{\partial z} \log{P_{\mathrm{even}}(z)}.
\eeq
Therefore the WKB solution can be expressed by only using the even part,
\beq
\label{eq:evensol}
 \psi_a(z) = \frac{1}{\sqrt{P_{\mathrm{even}}(z)}}\exp(\frac{i}{\eta}\int_{a}^{z}P_{\mathrm{even}}(z')dz').
\eeq
\par
$P_{\mathrm{even}}(z)dz$ in (\ref{eq:evensol}) can be regarded as a one-form on the Riemann surface $\Sigma$,
\begin{equation}
 \Sigma : y^2 = -Q_0(z).
\end{equation}
The branch points on $\Sigma$ are determined by the solutions to $Q_0\lr{z} = 0$. $\Sigma$ is a double covering of the complex plane with the branch points. $P_{\mathrm{even}}(z)dz$ admits the choice of the sign depending on the sheet of $\Sigma$. Each of the sign provides two linearly independent WKB solutions. We denote the two linearly independent solutions as,
\beq
\label{eq:pmsol}
 \psi_a^{\pm}(z) = \frac{1}{\sqrt{P_{\mathrm{even}}(z)}}\exp(\pm\frac{i}{\eta}\int_{a}^{z}P_{\mathrm{even}}(z')dz').
\eeq
A general solution to (\ref{eq:EWKBdeq}) is given by the linear combination of (\ref{eq:pmsol}). The WKB solutions (\ref{eq:pmsol}) can also be expanded as the series with respect to $\eta$ as follows,
\beq
 \psi_a^{\pm}(z) = \exp(\pm\frac{i}{\eta}\int_{a}^{z}p_0(z')dz')\sum_{n=0}^{\infty}\psi_{a, n}^{\pm}\lr{z}\eta^{n+\frac{1}{2}}.
\eeq
\par
The WKB solutions are asymptotic series, which converge only at $\eta =0$, and therefore need to be properly resumed. In the exact WKB analysis, we take the Borel resummation. The Borel resummation is defined as follows. Let us consider a asymptotic series,
\beq
 f = e^{-\frac{A}{\eta}}\sum_{n=0}^{\infty}a_n\eta^{n+\alpha}\ \ \lr{\alpha \notin \{-1, -2, -3, \dots\}}.
\eeq
We define the Borel transformation of $f$ as,
\beq
\label{eq:BT}
 \hat{f}\lr{\xi} = \sum_{n=0}^{\infty}\frac{a_n}{\Gamma\lr{n+\alpha}}\lr{\xi - A}^{n + \alpha -1}.
\eeq
If the series (\ref{eq:BT}) converges near $\xi = A$, then $f$ is said to be Borel summable \cite{cr1}. The Borel resummation of $f$ is defined as the Laplace integral of the Borel transformation,
\beq
 \mathcal{B}\lrk{f} = \int_{A}^{\infty}e^{-\frac{\xi}{\eta}}\hat{f}\lr{\xi}d\xi, 
\eeq
which is analytic on $\eta$ and has $f$ as the asymptotic expansion at $\eta = 0$. 
\par
The Borel resummed WKB solutions are local solutions at $z$. The global structure of the Borel resummed WKB solutions is determined by the Stokes lines, which are defined by the following equation,
\beq
\label{eq:SL}
 \Im{\lrk{i\int_{z_*}^{z}p_0\lr{z'}dz'}} = 0,
\eeq
where $z_*$ is a branch point on $\Sigma$. The collection of the Stokes lines is called the Stokes graph. The both ends of the stokes lines are connected to the branch points $z_*$ or the singular points $z=0, \infty$ of the differential equation (\ref{eq:EWKBdeq}). The Stokes lines split $\Sigma$ into some regions, which can be mapped into a triangulation of $\Sigma$ \cite{cr1, 0907.3987}. Each regions divided by the Stokes lines are called Stokes regions. 
\par
The Stokes lines have the orientation as follows:
\beq
\label{eq:Orientation}
\left\{
\begin{array}{l}
\Re\lrk{i\int_{z_*}^{z}p_0\lr{z'}dz'} > 0 \rightarrow \mathrm{positive\ orientation}, \\
\Re\lrk{i\int_{z_*}^{z}p_0\lr{z'}dz'} < 0 \rightarrow \mathrm{negative\ orientation}.
\end{array}
\right.
\eeq
The Borel resummed WKB solutions provide local and analytic solutions to (\ref{eq:EWKBdeq}) in a Stokes region.  When we analytically continue the solutions from a Stokes region to another, the solutions are changed discontinuously. This phenomenon is called the Stokes phenomenon of the WKB solutions. In the next section, we will study a connection problem of the Borel resummed WKB solutions for (\ref{eq:EWKBdeq}) to obtain the spectra of $\omega$.

\section{Connection problem and quasi-normal modes}
\label{sec:CPQNM}

\subsection{Connection formula for the WKB solutions}
\label{subsec:ACWKB}
We will consider a connection problem for the solutions to (\ref{eq:EWKBdeq}) from $z \rightarrow +\infty$ to the origin. In the following, we fix the parameters at certain values $\omega L = \frac{13}{10} - i\frac{1}{16}$ and $l = 0$ to illustrate the problem.

\begin{figure}[htbp]
  \begin{center}
    \includegraphics[clip, width=8.0cm]{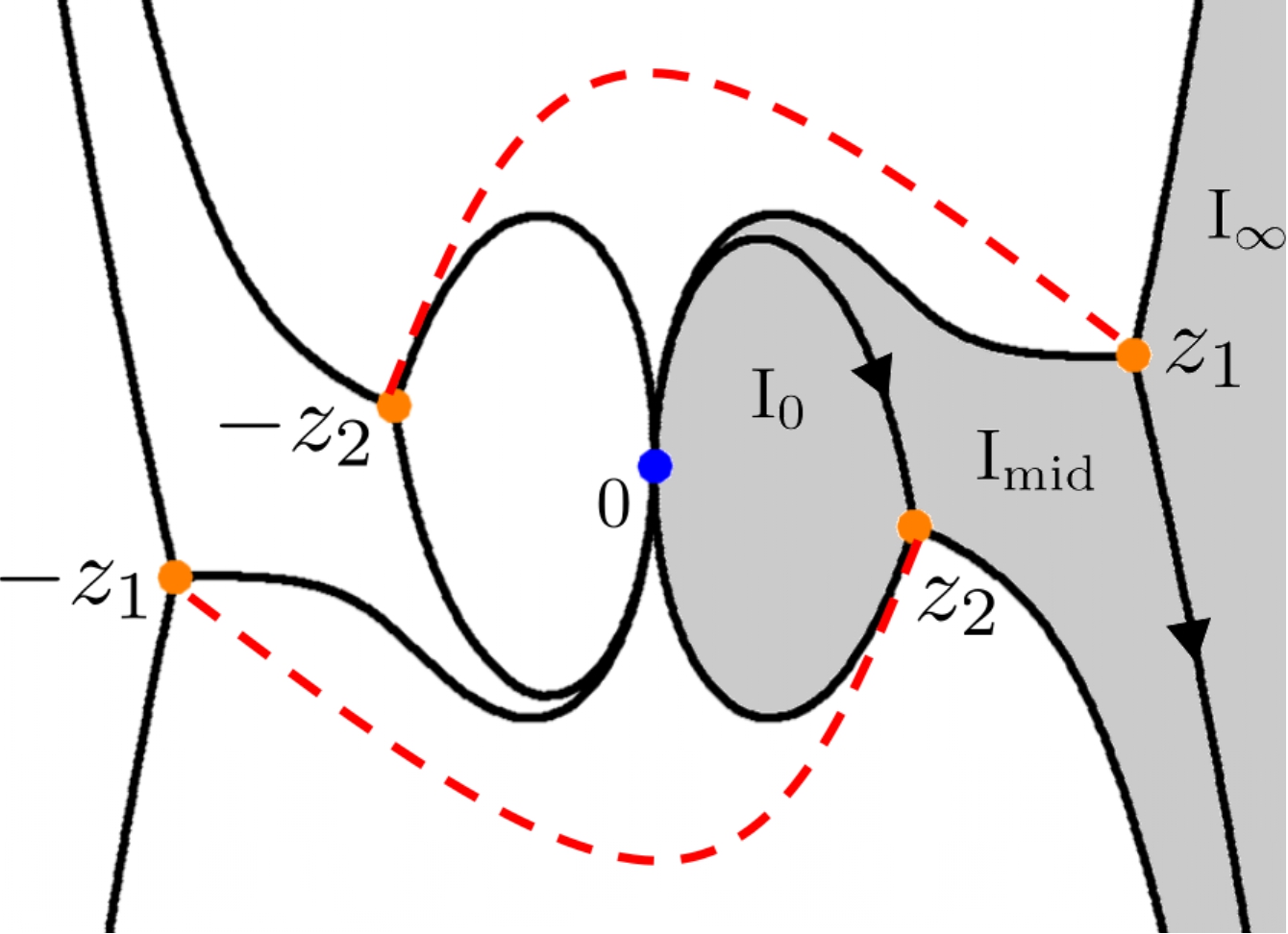}
    \caption{The Stokes lines (the black lines) at $\omega L = \frac{13}{10} - \frac{i}{16}$, $l = 0$. The sheet of $\Sigma$ is taken so that $P_{\mathrm{even}}(z)dz$ has the positive sign. The orange points $\pm z_1, \pm z_2$ are the branch points and the red dashed lines are the branch cuts. The blue point indicates the origin, which is an irregular singularity of the differential equation (\ref{eq:EWKBdeq}). The black arrows on the stokes lines indicate the orientations of them (for the negative orientation, the arrow points to the branch point along the Stokes line, and for the positive orientation, the arrow points to the singular point.). The stokes lines going out of the frame flow at the infinity, which is another irregular singularity of (\ref{eq:EWKBdeq}).}
    \label{fig:CP}
  \end{center}
\end{figure}

\par
The Stokes graph is drawn in Fig.\ref{fig:CP}, where the Stokes regions are labeled by $\mathrm{I}_{\infty}$, $\mathrm{I}_\mathrm{mid}$ and $\mathrm{I}_0$. We consider the connection problem of the Borel resummed WKB solutions starting from $\mathrm{I}_{\infty}$ to $\mathrm{I}_\mathrm{mid}$ and $\mathrm{I}_\mathrm{mid}$ to $\mathrm{I}_0$ based on \cite{DP, cr1, 2008.00379}. In the analytic continuation from $\mathrm{I}_{\infty}$ to $\mathrm{I}_\mathrm{mid}$, the WKB solutions cross over a Stokes line emanating from $z_1$ and oriented to the positive direction clockwisely viewing from $z_1$. Then the Borel resummed WKB solutions in the region $\mathrm{I}_{\infty}$ and in the region $\mathrm{I}_{\mathrm{mid}}$ are shown to connect by the following formula \cite{DP, cr1, 2008.00379},
\beq
\label{eq:infmid}
\left( \begin{array}{r}
\mathcal{B}\lrk{\psi_{z_1, \mathrm{I_{\infty}}}^{+}}(z) \\ \mathcal{B}\lrk{\psi_{z_1, \mathrm{I_{\infty}}}^{-}}(z)
\end{array} \right) =  
\left( \begin{array}{rr}
1 & -i \\ 0 & 1
\end{array} \right) \left( \begin{array}{r}
\mathcal{B}\lrk{\psi_{z_1, \mathrm{I}_{\mathrm{mid}}}^{+}}(z) \\ \mathcal{B}\lrk{\psi_{z_1, \mathrm{I}_{\mathrm{mid}}}^{-}}(z)
\end{array} \right),
\eeq
where we denoted the solutions in $\mathrm{I}_{\infty}$ as $\psi_{z_1, \mathrm{I_{\infty}}}^{\pm}(z)$, and the solutions in $\mathrm{I}_{\mathrm{mid}}$ in the same manner. 
Next, in the analytic continuation from $\mathrm{I}_\mathrm{mid}$ to $\mathrm{I}_{0}$, the solutions cross over a Stokes line emanating from $z_2$ and oriented to the negative direction anti-clockwisely viewing from $z_2$. In order to apply the connection formula, we need to change the normalization of the solutions \cite{2008.00379},
\beq
\label{eq:midnor}
\left( \begin{array}{r}
\mathcal{B}\lrk{\psi_{z_1, \mathrm{I}_{\mathrm{mid}}}^{+}}(z) \\ \mathcal{B}\lrk{\psi_{z_1, \mathrm{I}_{\mathrm{mid}}}^{-}}(z)
\end{array} \right) =  
\left( \begin{array}{rr}
e^{-\frac{i}{\eta}\mathcal{B}\lrk{\int_{z_2}^{z_1}P_{\mathrm{even}}(z)dz}} & 0 \\ 0 & e^{\frac{i}{\eta}\mathcal{B}\lrk{\int_{z_2}^{z_1}P_{\mathrm{even}}(z)dz}}
\end{array} \right) \left( \begin{array}{r}
\mathcal{B}\lrk{\psi_{z_2, \mathrm{I}_{\mathrm{mid}}}^{+}}(z) \\ \mathcal{B}\lrk{\psi_{z_2, \mathrm{I}_{\mathrm{mid}}}^{-}}(z)
\end{array} \right).
\eeq
The solutions in $\mathrm{I}_{\mathrm{mid}}$ and in $\mathrm{I}_{0}$ normalized at $z_2$
are connected as follows \cite{DP, cr1, 2008.00379},
\beq
\label{eq:mid0}
\left( \begin{array}{r}
\mathcal{B}\lrk{\psi_{z_2, \mathrm{I}_{\mathrm{mid}}}^{+}}(z) \\ \mathcal{B}\lrk{\psi_{z_2, \mathrm{I}_{\mathrm{mid}}}^{-}}(z)
\end{array} \right) =  
\left( \begin{array}{rr}
1 & 0 \\ i & 1
\end{array} \right) \left( \begin{array}{r}
\mathcal{B}\lrk{\psi_{z_2, \mathrm{I}_{0}}^{+}}(z) \\ \mathcal{B}\lrk{\psi_{z_2, \mathrm{I}_{0}}^{-}}(z)
\end{array} \right).
\eeq
By multiplying (\ref{eq:infmid})$\sim$(\ref{eq:mid0}) and finally changing the normalization of $\psi_{z_2, \mathrm{I}_{0}}^{\pm}$ in the r.h.s. of (\ref{eq:mid0}) to $z_1$ in the same way as (\ref{eq:midnor}), we obtain the connection formula from $\mathrm{I}_{\infty}$ to $\mathrm{I}_{0}$,
\beq
\label{eq:FCF}
\left( \begin{array}{r}
\mathcal{B}\lrk{\psi_{z_1, \mathrm{I_{\infty}}}^{+}}(z) \\ \mathcal{B}\lrk{\psi_{z_1, \mathrm{I_{\infty}}}^{-}}(z)
\end{array} \right) =  
\left( \begin{array}{rr}
1 + e^{\frac{i}{\eta}\mathcal{B}\lrk{2\int_{z_2}^{z_1}P_{\mathrm{even}}(z)dz}} & -i \\ ie^{\frac{i}{\eta}\mathcal{B}\lrk{2\int_{z_2}^{z_1}P_{\mathrm{even}}(z)dz}} & 1
\end{array} \right) \left( \begin{array}{r}
\mathcal{B}\lrk{\psi_{z_1, \mathrm{I_{0}}}^{+}}(z) \\ \mathcal{B}\lrk{\psi_{z_1, \mathrm{I_{0}}}^{-}}(z)
\end{array} \right).
\eeq
\par
\begin{figure}[htbp]
  \begin{center}
    \includegraphics[clip, width=12.9cm]{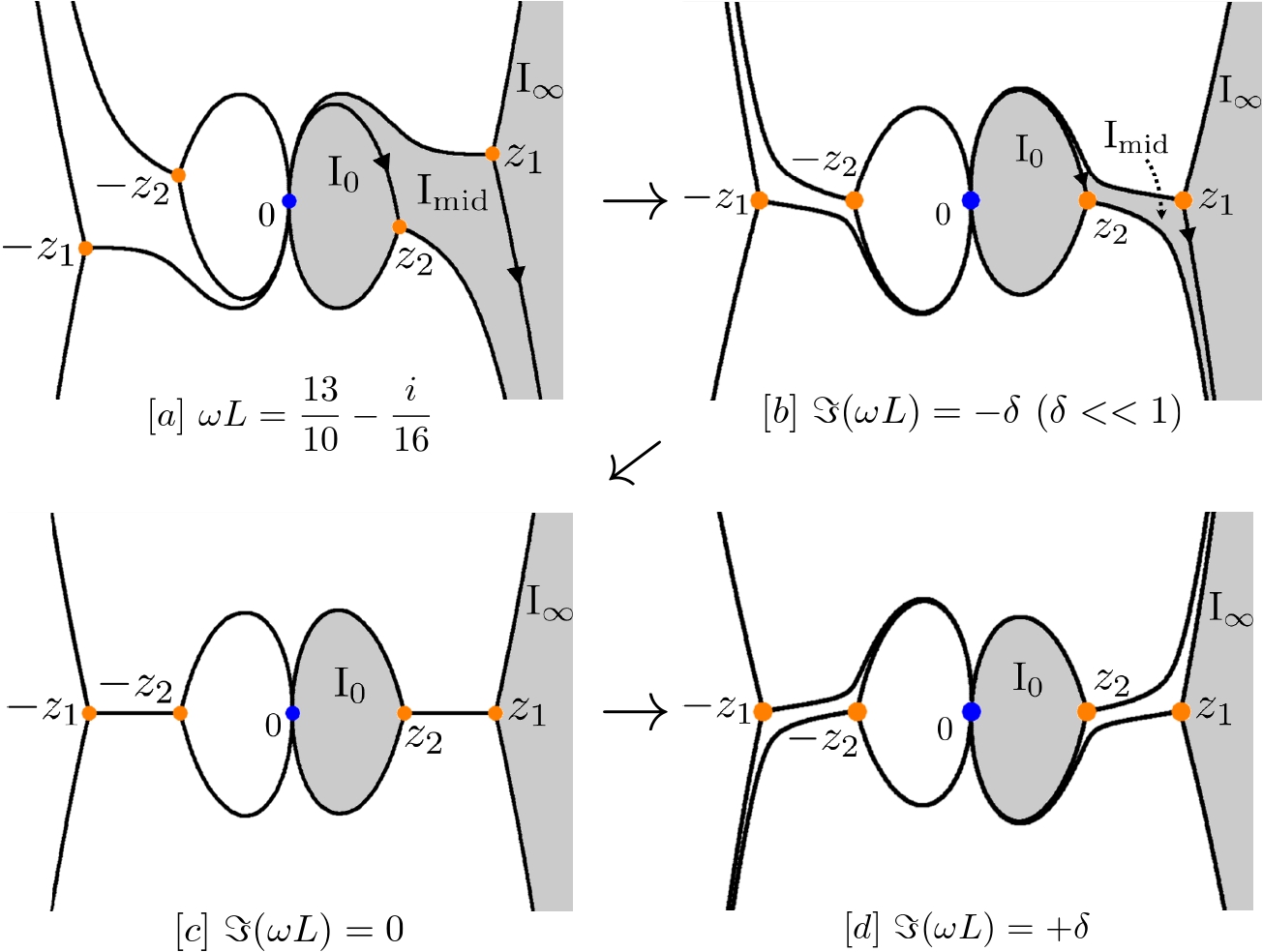}
    \caption{The deformation of the Stokes graph.}
    \label{fig:varySG}
  \end{center}
\end{figure}

In the above derivation, we fixed the value of $\omega L$ and $l$. Therefore we need to discuss how the parameter regions our derivation is valid. First, let us consider what happen if we vary $\omega L$ from $\omega L = 13/10 - i/16$ to $\omega L = 13/10 + i\delta \ (\delta << 1)$. Fig.\ref{fig:varySG} shows the transition of the Stokes graph. The graph [a] is the same as Fig.\ref{fig:CP}. In the graph [b], the Stokes region $\mathrm{I}_{\mathrm{mid}}$ narrows but survives. For this case, we obtain same connection formula as (\ref{eq:FCF}) because each connection formula for the Borel resummed WKB solutions that we have used to compute (\ref{eq:FCF}) only depends on which two adjacent Stokes regions we connect and the orientation of the Stokes line sandwiched by them. This example indicates that we obtain same connection formula (\ref{eq:FCF}) if the Stokes graph has the Stokes regions $\mathrm{I}_{0}, \mathrm{I}_{\mathrm{mid}}, \mathrm{I}_{\infty}$ (regardless of their area). Conversely, we do not obtain the connection formula (\ref{eq:FCF}) in the case of the graph [c]. Two Stokes lines emanating from $z_1$ and $z_2$ collide and become one Stokes line connecting $z_1$ and $z_2$. And then the Stokes region $\mathrm{I}_{\mathrm{mid}}$ is completely crushed. Continuing to increase the imaginary part of $\omega L$, we obtain the graph [d]. The two Stokes lines collided in [c] are separated again, but their position is inverted and there is no longer the Stokes region $\mathrm{I}_{\mathrm{mid}}$. This phenomenon is called the "flip" of the Stokes graph \cite{cr1}.

\begin{figure}[htbp]
  \begin{center}
   \hspace*{-1.0in}
    \includegraphics[clip, width=8.0cm]{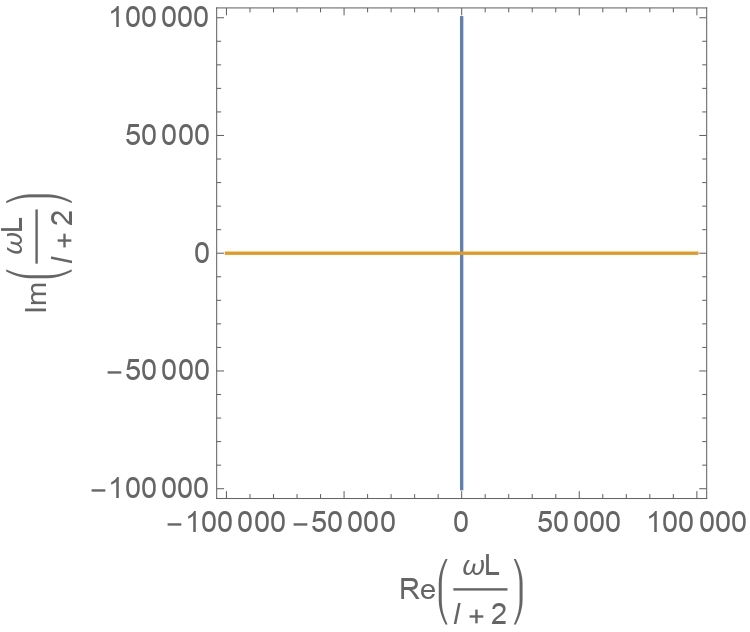}
    \caption{The orange line indicates the values of $\frac{\omega L}{l+2}$ satisfying (\ref{eq:bcond2}) and the blue line indicates the values satisfying (\ref{eq:acond2}).}
    \label{fig:abRegion}
  \end{center}
\end{figure}

As the example of Fig.\ref{fig:varySG}, if we continuously change the parameters $\omega L$ and $l$, the Stokes graph (and the triangulation of $\Sigma$ mapped from the Stokes graph) may be discontinuously changed by the flip. Then the derivation of (\ref{eq:FCF}) is no longer valid. The sign of the flip of the Stokes graph is to appear the Stokes line which connects two branch points as the graph [c] in Fig.\ref{fig:varySG}. The set of two branch points which are adjacent to a same Stokes region may be connected by the Stokes line. In the case of Fig.\ref{fig:CP}, the branch points $z_1$ and $-z_2$, $-z_1$ and $z_2$, $z_1$ and $z_2$, $-z_1$ and $-z_2$ may be  connected each other. By definition of the Stokes line (\ref{eq:SL}), the conditions that these sets of the branch points are connected are the following equations,
\beq
\label{eq:bcond}
 \Im\lrk{2i\int_{z_2}^{z_1}p_0\lr{z}dz} = \Im\lrk{2i\int_{-z_1}^{-z_2}p_0\lr{z}dz} = 0,
\eeq
\beq
\label{eq:acond}
 \Im\lrk{2i\int_{z_1}^{-z_2}p_0\lr{z}dz} = \Im\lrk{2i\int_{-z_2}^{z_1}p_0\lr{z}dz} = 0,
\eeq
where we used the condition $p_0(-z) = p_0(z)$. Each of the integrals can be expressed by the hypergeometric functions,
\beq
\label{eq:b0}
 2i\int_{z_2}^{z_1}p_0\lr{z}dz = \pi \lr{1-\frac{\lr{l+2}^2}{2\omega^2L^2}}\omega L\ _2F_1\lrk{\frac{1}{2}, \frac{1}{2}, 2, \frac{1}{2}\lr{1-\frac{\lr{l+2}^2}{2\omega^2L^2}}},
\eeq
\beq
\label{eq:a0}
 2i\int_{z_1}^{-z_2}p_0\lr{z}dz = i\pi \lr{1+\frac{\lr{l+2}^2}{2\omega^2L^2}}\omega L\ _2F_1\lrk{\frac{1}{2}, \frac{1}{2}, 2, \frac{1}{2}\lr{1+\frac{\lr{l+2}^2}{2\omega^2L^2}}}.
\eeq
(\ref{eq:b0}) and (\ref{eq:a0}) are local expressions but we can also compute each integrals for other parameter regions by analytically continuing the hypergeometric functions. Dividing (\ref{eq:bcond}) and (\ref{eq:acond}) by $(l+2)$ and substituting (\ref{eq:b0}) and (\ref{eq:a0}), we obtain the following conditions,
\beq
\label{eq:bcond2}
 \Im\lrk{\pi \lr{1-\frac{\lr{l+2}^2}{2\omega^2L^2}}\frac{\omega L}{(l+2)}\ _2F_1\lrk{\frac{1}{2}, \frac{1}{2}, 2, \frac{1}{2}\lr{1-\frac{\lr{l+2}^2}{2\omega^2L^2}}}} = 0,
\eeq
\beq
\label{eq:acond2}
 \Im\lrk{i\pi \lr{1+\frac{\lr{l+2}^2}{2\omega^2L^2}}\frac{\omega L}{(l+2)}\ _2F_1\lrk{\frac{1}{2}, \frac{1}{2}, 2, \frac{1}{2}\lr{1+\frac{\lr{l+2}^2}{2\omega^2L^2}}}} = 0.
\eeq
In these expressions, there is only one parameter $\omega L / (l + 2)$. Therefore we only need to check which value of $\omega L / (l + 2)$ satisfies (\ref{eq:bcond2}), (\ref{eq:acond2}). We can immediately find that (\ref{eq:bcond2}) is satisfied by $\omega L / (l + 2) \in \mathbb{R}$ and (\ref{eq:acond2}) is satisfied by $\omega L / (l + 2) \in i\mathbb{R}$. Fig.\ref{fig:abRegion} shows the points satisfying (\ref{eq:bcond2}) or (\ref{eq:acond2}). In Fig.\ref{fig:abRegion}, there are no points in the region $\Re{(\omega L / (l+2))} > 0$ and $\Im{(\omega L / (l+2))} < 0$, which includes $\omega L = \frac{13}{10} - i\frac{1}{16}$, $l=0$. Therefore the derivation of (\ref{eq:FCF}) is valid for $\Re{(\omega L / (l+2))} > 0$ and $\Im{(\omega L / (l+2))} < 0$ region, which includes the QNMs, at least in the range depicted in Fig.\ref{fig:abRegion}. We can also check arbitrary value of $\omega L / (l+2)$ by using (\ref{eq:bcond2}) and (\ref{eq:acond2}).

\subsection{Boundary conditions and QNMs}
\label{subsec:BCQNM}
The quasi-normal modes for the D3-branes are the discrete set of $\omega$ at which the wavefunction satisfies the following two boundary conditions \cite{2105.04245},
\par
(i) Only outgoing wave exists at $z \rightarrow +\infty$.
\par
(ii) Only ingoing wave exists at $z \rightarrow 0$.
\\
The boundary condition at $+\infty$ is satisfied by $\mathcal{B}\lrk{\psi_{z_1, \mathrm{I_{\infty}}}^{+}}(z)$. From the connection formula (\ref{eq:FCF}), $\mathcal{B}\lrk{\psi_{z_1, \mathrm{I_{\infty}}}^{+}}(z)$ is analytically connected to the origin as,
\beq
\label{eq:BCCF}
\mathcal{B}\lrk{\psi_{z_1, \mathrm{I_{\infty}}}^{+}}(z) = \lr{1 + e^{\frac{i}{\eta}\mathcal{B}\lrk{2\int_{z_2}^{z_1}P_{\mathrm{even}}(z)dz}}}\mathcal{B}\lrk{\psi_{z_1, \mathrm{I_{0}}}^{+}}(z) -i\mathcal{B}\lrk{\psi_{z_1, \mathrm{I_{0}}}^{-}}(z).
\eeq 
Because the boundary condition at $0$ is satisfied by $\mathcal{B}\lrk{\psi_{z_1, \mathrm{I_{0}}}^{-}}(z)$, (\ref{eq:BCCF}) needs to satisfy the following equation,
\beq
\label{eq:BSBS}
 1 + e^{\frac{i}{\eta}\mathcal{B}\lrk{2\int_{z_2}^{z_1}P_{\mathrm{even}}(z)dz}} = 0.
\eeq
Substituting $\eta = 1$ and rearranging the equation, we obtain the condition for the QNMs,
\beq
\label{eq:BSBS2}
 \mathcal{B}\lrk{2\int_{z_2}^{z_1}P_{\mathrm{even}}(z)dz} = 2\pi\lr{n + \frac{1}{2}}, \ \ \lr{n \in \mathbb{Z}_{\geq 0}}.
\eeq
Without considering convergence of the asymptotic series, the asymptotic expansion of the Borel resummation provides the all-order Bohr-Sommerfeld condition,
\beq
\label{eq:AOBS}
 2\int_{z_2}^{z_1}P_{\mathrm{even}}(z)dz = 2\pi\lr{n + \frac{1}{2}}.
\eeq
\par
In \cite{2105.04245, 2109.09804}, the r.h.s. of (\ref{eq:AOBS}) is $2\pi n$ instead of $2\pi\lr{n + \frac{1}{2}}$. This difference is caused by the difference of the definition for $P_{\mathrm{even}}(z)$. In this paper, we define $P_{\mathrm{even}}(z)$ as a power series with respect to $\eta$. On the other hand, in \cite{2105.04245, 2109.09804}, $P_{\mathrm{even}}(z)$ is a power series with respect to $1 / \omega L$. Therefore $P_{\mathrm{even}}(z)$ in this paper is not the same ones in \cite{2105.04245, 2109.09804}
\par
I would like to make a comment for (\ref{eq:BSBS2}) and (\ref{eq:AOBS}) from the point of view of the resurgence. In the resurgent quantum mechanics, we can derive the energy quantization conditions by solving the connection problem as we have done. Then the energy quantization conditions often have exponentially small terms, which are connected to the non-perturbative contributions to the asymptotic expansion of the energy \cite{Esemi}. On the other hand, the condition for the QNMs (\ref{eq:BSBS2}) and (\ref{eq:AOBS}) has no exponentially small terms. Therefore our results indicate there are no non-perturbative contributions to the asymptotic expansion of the QNMs.
\par
We will check the consistency of the condition (\ref{eq:BSBS2}). To calculate the Borel resummation of $2\int_{z_2}^{z_1}P_{\mathrm{even}}(z)dz$, we use the Borel-Pad\'{e} approximation,
\beq
 \mathcal{B}\lrk{2\int_{z_2}^{z_1}P_{\mathrm{even}}(z)dz} \sim \int_{0}^{\infty}e^{-\xi}\lrk{N/N}d\xi,
\eeq
where $\lrk{N/N}$ indicates the diagonal Pad\'{e} approximation of order $N$ for the Borel transformation of $2\int_{z_2}^{z_1}P_{\mathrm{even}}(z)dz$. We will use $N=4$ case with 8th-order coefficients of $2\int_{z_2}^{z_1}P_{\mathrm{even}}(z)dz$. The higher-order coefficients of $2\int_{z_2}^{z_1}P_{\mathrm{even}}(z)dz$ can be computed by the differential operator technique (e.g. \cite{0910.5670, 1504.08324, 1006.5185, 2002.06829}),
\beq
 \int_{z_2}^{z_1}p_{n}(z)dz = \int_{z_2}^{z_1}\mathcal{O}_n p_{0}(z)dz = \mathcal{O}_n\int_{z_2}^{z_1}p_{0}(z)dz,
\eeq
where $\mathcal{O}_n$ is a differential operator with respect to $u = \lr{l+2}^2$. Up to 8th-order, the differential operators are given as follows \footnote{The differential operators (\ref{eq:diffop}) are the same ones for the Mathieu equation computed in e.g. \cite{0910.5670, 1504.08324, 1006.5185, 2002.06829} because (\ref{eq:EWKBdeq}) agrees with the Mathieu equation under the change of variable $z = e^{iq}$ and then the one-form $P_{\mathrm{even}}(z)dz$ is invariant (Proposition 2.7. (b) in \cite{cr1}).}:
\beq
\label{eq:diffop}
\begin{split}
 &\mathcal{O}_2 = \frac{u}{3}\frac{\partial^2}{\partial u^2} + \frac{1}{6}\frac{\partial}{\partial u}, \\
 &\mathcal{O}_4 = \frac{1}{2^3}\lr{\frac{28 u^2}{45}\frac{\partial^4}{\partial u^4} + \frac{8u}{3}\frac{\partial^3}{\partial u^3} + \frac{5}{3}\frac{\partial^2}{\partial u^2}}, \\
 &\mathcal{O}_6 = \frac{1}{2^3}\lr{\frac{124 u^3}{945}\frac{\partial^6}{\partial u^6} + \frac{158 u^2}{105}\frac{\partial^5}{\partial u^5} + \frac{153u}{35}\frac{\partial^4}{\partial u^4} + \frac{41}{14}\frac{\partial^3}{\partial u^3}}, \\
 &\mathcal{O}_8 = \frac{127 u^4}{2^3\times 4725}\frac{\partial^8}{\partial u^8} + \frac{13u^3}{175}\frac{\partial^7}{\partial u^7} + \frac{517u^2}{2^4\times 63}\frac{\partial^6}{\partial u^6} + \frac{9539u}{2^3\times 945}\frac{\partial^5}{\partial u^5} + \frac{15229}{2^7\times 135}\frac{\partial^4}{\partial u^4}.
\end{split}
\eeq
\begin{table}[tp]
\begin{center}
\hspace*{-0.4in} 
\begin{tabular}{cccc} \hline
$l$ & $\omega L$ (Leaver's method) & $2\pi\lr{0 + \frac{1}{2}}$ & $\mathcal{B}\lrk{2\int_{z_2}^{z_1}P_{\mathrm{even}}(z)dz}$  \\ \hline
0 & $1.369686 - 0.503996i$ & $\underline{3.141592}$ & $\underline{3.14159}3 - 1.479873\times 10^{-6}i$ \\
1 & $2.091757 - 0.501753i$ & $\underline{3.141592}$ & $\underline{3.141592} - 4.412743\times 10^{-8}i$ \\
2 & $2.806287 - 0.500981i$ & $\underline{3.141592}$ & $\underline{3.141592} - 3.550133\times 10^{-9}i$ \\ \hline
\end{tabular}
\caption{Numerical check of (\ref{eq:BSBS2}) with $n = 0$ case.}
\label{tab:CCBPn0}
\end{center}
\end{table}
In Table.\ref{tab:CCBPn0}, we show the quasi-normal modes with $n=0$ case obtained by the Leaver's method \cite{2105.04245, Leaver} and the values of $\mathcal{B}\lrk{2\int_{z_2}^{z_1}P_{\mathrm{even}}(z)dz}$ at that time.
\par
We can also obtain an analytic expression of the QNMs. The integral $2\int_{z_2}^{z_1}p_{n}(z)dz$ can be expanded at $\omega L = \frac{l+2}{\sqrt{2}}$. For example, we obtain
\beq
\begin{split}
 &2\int_{z_2}^{z_1}p_{0}(z)dz = -2 i \pi  \left(\omega L-\frac{l+2}{\sqrt{2}}\right)+\frac{3 i \pi  \left(\omega L-\frac{l+2}{\sqrt{2}}\right)^2}{2 \sqrt{2} (l+2)}+\dots \ ,\\
 &2\int_{z_2}^{z_1}p_{2}(z)dz = \frac{i \pi }{8 \sqrt{2} (l+2)} -\frac{3 i \pi  \left(\omega L-\frac{l+2}{\sqrt{2}}\right)}{64 (l+2)^2} - \frac{i \pi  \left(\omega L-\frac{l+2}{\sqrt{2}}\right)^2}{256 \sqrt{2} (l+2)^3} + \dots \ ,\\
 &2\int_{z_2}^{z_1}p_{4}(z)dz = -\frac{17 i \pi }{2048 \sqrt{2} (l+2)^3} + \frac{95 i \pi  \left(\omega L-\frac{l+2}{\sqrt{2}}\right)}{32768 (l+2)^4} + \frac{1959 i \pi  \left(\omega L-\frac{l+2}{\sqrt{2}}\right)^2}{131072 \sqrt{2} (l+2)^5} + \dots \ .
\end{split}
\eeq
Then the condition (\ref{eq:AOBS}) can be expressed as 
\beq
\label{eq:EXPBS}
2\pi\lr{n + \frac{1}{2}} = 2\int_{z_2}^{z_1}p_{0}(z)dz + 2\int_{z_2}^{z_1}p_{2}(z)dz + 2\int_{z_2}^{z_1}p_{4}(z)dz + \dots = \sum_{m=0}^{\infty}c_m\lr{\omega L - \frac{l+2}{\sqrt{2}}}^m.
\eeq
(\ref{eq:EXPBS}) can be inverted by using the Lagrange inversion theorem. Then we obtain an analytic form of the QNMs,
\beq
\label{eq:ANQNM}
\hspace*{-0.4in}
 \omega_n L = \frac{l+2}{\sqrt{2}} + \frac{2\pi\lr{n + \frac{1}{2}}-c_0}{c_1} - \frac{c_2 (2\pi\lr{n + \frac{1}{2}}-c_0)^2}{c_1^3} + \dots
\eeq
In Table.\ref{tab:QNMn0} and \ref{tab:QNMn1}, we numerically compare the Leaver's method and (\ref{eq:ANQNM}), where we have used up to 16-th order of (\ref{eq:ANQNM}).

\subsubsection*{Large $n$ limit}
\par
In the large $n$ limit (which corresponds to $|\omega L| \gg (l+2)$ limit), (\ref{eq:b0}) and (\ref{eq:a0}) become
\beq
 2i\int_{z_2}^{z_1}p_0\lr{z}dz = \pi \omega L\ _2F_1\lrk{\frac{1}{2}, \frac{1}{2}, 2, \frac{1}{2}},
\eeq
\beq
 2i\int_{z_1}^{-z_2}p_0\lr{z}dz = i\pi \omega L\ _2F_1\lrk{\frac{1}{2}, \frac{1}{2}, 2, \frac{1}{2}},
\eeq
and then (\ref{eq:bcond}), (\ref{eq:acond}) are equivalent to
\beq
 \Im\lrk{\omega} = 0,
\eeq
\beq
 \Im\lrk{i \omega} = \Re\lrk{\omega} = 0.
\eeq
Therefore, in the large $n$ limit, the QNM conditions (\ref{eq:BSBS2}) and (\ref{eq:AOBS}) are valid for the region $\omega = \omega_{\mathrm{R}} - i\omega_{\mathrm{I}}$ with $\omega_{\mathrm{R}}, \omega_{\mathrm{I}} \in \mathbb{R}_{> 0}$. 
\par
In the large $n$ limit, the leading order approximation (or the WKB approximation) is valid. Then the QNM condition (\ref{eq:AOBS}) becomes
\beq
 2\int_{z_2}^{z_1}p_0\lr{z}dz = -\pi \omega L \ _2F_1\lrk{\frac{1}{2}, \frac{1}{2}, 2, \frac{1}{2}} = 2i\pi \lr{n + \frac{1}{2}}.
\eeq
This reads the QNMs at large $n$ is pure imaginary,
\beq
\label{eq:largen}
 \omega L = -i\frac{2\lr{n + \frac{1}{2}}}{\ _2F_1\lrk{\frac{1}{2}, \frac{1}{2}, 2, \frac{1}{2}}}.
\eeq
In some example, it is confirmed that the real part of the QNMs at large $n$ limit is proportional to the entropy of the black hole (e.g. 4d Schwarzschild BH \cite{gr-qc/0212096}, 4d Schwarzschild BH and 4d Reissner-Nordstr$\rm{\ddot{o}}$m BH \cite{gr-qc/0307020}, d$\geq$4 Schwarzschild BH and 4d Reissner-Nordstr$\rm{\ddot{o}}$m BH \cite{gr-qc/0301173}). It is also true for the present case (\ref{eq:largen}) because the entropy of the D3-brane metric is zero \cite{hep-th/9602135}.

\begin{table}[tp]
\begin{center}
\hspace*{-0.4in} 
\begin{tabular}{ccc} \hline
$l$ & Leaver's method & (\ref{eq:ANQNM})  \\ \hline
0 & $\underline{1.36968}6 - \underline{0.50399}6i$ & $\underline{1.36968}2 - \underline{0.50399}3i$ \\
1 & $\underline{2.09175}7 - \underline{0.501753}i$ & $\underline{2.09175}6 - \underline{0.501753}i$ \\
2 & $\underline{2.806287} - \underline{0.500981}i$ & $\underline{2.806287} - \underline{0.500981}i$ \\ \hline
\end{tabular}
\caption{$n = 0$ case.}
\label{tab:QNMn0}
\end{center}
\end{table}

\begin{table}[tp]
\begin{center}
\hspace*{-0.4in} 
\begin{tabular}{ccc} \hline
$l$ & Leaver's method & (\ref{eq:ANQNM})  \\ \hline
0 & $\underline{0.682}086 - \underline{1.559}933i$ & $\underline{0.682}296 - \underline{1.559}289i$ \\
1 & $\underline{1.704812} - \underline{1.5186}80i$ & $\underline{1.704812} - \underline{1.5186}79i$ \\
2 & $\underline{2.528152} - \underline{1.509655}i$ & $\underline{2.528152} - \underline{1.509655}i$ \\ \hline
\end{tabular}
\caption{$n = 1$ case.}
\label{tab:QNMn1}
\end{center}
\end{table}

\section{Summary and discussions}
\label{sec:CaD}
In this paper, we applied the Exact WKB analysis to the QNMs eigenvalue problem of a massless scalar perturbation to the D3-brane metric. The Exact WKB analysis provides a solution to the connection problem. Combining with the boundary conditions, we obtained the Borel-resummed Bohr-Sommerfeld condition (\ref{eq:BSBS2}), which has the all-order Bohr-Sommerfeld condition (\ref{eq:AOBS}) as the asymptotic expansion. We numerically checked the condition (\ref{eq:BSBS2}) 
by using the Borel-Pad\'{e} approximation. We also presented an analytic form of the QNMs (\ref{eq:ANQNM}) and the large n limit (\ref{eq:largen}).
\par
I would like to comment for the technical advantage of the Exact WKB analysis. The equation (\ref{eq:EWKBdeq}) corresponds to the quantum mechanical system that there is one potential barrier in the radial direction (see e.g. \cite{classicalWKB}). In this case, we can expect that the QNMs can be computed by the all-order Bohr-Sommerfeld condition for the potential barrier as (\ref{eq:AOBS}). It is also the case for the all-order Bohr-Sommerfeld conjecture previously studied in \cite{2006.06111, 2105.04245, 2109.09804}. But if the situation we are considering has more than one potential barrier, then not only we cannot determine which potential barrier produces the all-order Bohr-Sommerfeld condition for the QNMs but also there is no guarantee that the all-order Bohr-Sommerfeld condition is valid. The Exact WKB analysis may provides a systematic way to derive the condition satisfied by the QNMs even for these case.
\par
Other direction of the future work is the application to the reflection and the absorption probability \cite{hep-th/9805140}. The connection problem of the in-going WKB solutions at $z \rightarrow +\infty$ provides the reflection and the penetration wave to the potential barrier. The ratios of each wave provide the reflection and the absorption probability. It is also interesting to discuss the meaning of the Bohr-Sommerfeld condition in the context of the D-brane world volume picture \cite{hep-th/0208063}. 

\section*{Acknowledgements}
We would like to thank Katsushi Ito and Shota fujiwara for valuable discussions and comments. This work is supported in part by JSPS Research Fellowship 22J15182 for Young Scientists, from Japan Society for the Promotion of Science (JSPS).

\end{document}